\documentclass[11pt,nofootinbib,showpacs]{revtex4}
\usepackage{amsfonts,amssymb,amsmath,graphicx}
\def\[{\left[}
\def\]{\right]}
\def\({\left(}
\def\){\right)}

\begin{document}

\title{The dynamics of the flat anisotropic models in the Lovelock gravity. \linebreak I: The even-dimensional case}

\author{S.A. Pavluchenko}
\affiliation{Special Astrophysical Observatory, Russian Academy of
Sciences, Nizhnij Arkhyz, 369167 Russia}

\begin{abstract}
In this article we give a full description of the dynamics of the
flat anisotropic (4+1)-dimensional cosmological model in the
presence of both Gauss-Bonnet and Einstein contributions. This is
the first complete description of this model with both terms taken
into account. Our data is obtained using the numerical analysis,
though, we use analytics to explain some features of the results
obtained, and the same analytics could be applied to
higher-dimensional models in higher-order Lovelock corrections.
Firstly, we investigate the vacuum model and give a description of
all regimes; then, we add a matter source in the form of a perfect
fluid and study the influence the matter exerts upon the dynamics.
Thus, we give a description of matter regimes as well.
Additionally, we demonstrate that the presence of matter not only
``improves'' the situation with a smooth transition between the
standard singularity and the Kasner regime, but also brings
additional regimes and even partially ``erases'' the boundaries
between different regimes inside the same triplet. Finally, we
discuss the numerical and analytical results obtained and their
generalization to the higher-order models.
\end{abstract}


\maketitle

\section{Introduction}

The idea of extra dimensions traces back to the beginning of last
century, to the papers by Nordstr\"om~\cite{Nord1914},
Kaluza~\cite{Kaluza} and Klein~\cite{Klein}. At that time such
ideas were considered as mere mathematical speculations, but with
time they have come to hold a firm place in the minds of
scientists. The 70s gave rise to an interest in the
extradimensional theories due to the development of superstring
and supergravity theories (see, e.g., \cite{super}). Later in the
90s the interest has only increased owing to the possibility to
solve the hierarchy problem~\cite{hierarchy}, lower the grand
unification scale in the M-theory~\cite{M-GUT} and in many others
(see, e.g.,\cite{rec-rev} for recent reviews).

Dealing with extra dimensions in the cosmological context, one
usually uses a modified gravity (see the forementioned reviews).
One of modifications of this kind is the Lovelock
gravity~\cite{Lovelock}. This theory is equivalent to Einstein's
theory in (3+1), but starting from (4+1) it gives rise to
higher-order (in powers of curvature) corrections. The first of
these corrections is known as the Gauss-Bonnet term, first found
by Lanczos~\cite{Lanczos} (therefore it is sometimes referred to
as the Lanczos term). In the cosmological context the Lovelock
(and Gauss-Bonnet in particular) gravity was intensively studied
over the past two decades~\cite{studied}.

In this paper we investigate the dynamics of the flat anisotropic
Universe in Lovelock gravity. Since we work in a (4+1)-dimensional
case, Lovelock contribution reduces to the Gauss-Bonnet term. We
took both the General Relativity (GR) and Gauss-Bonnet (GB)
contributions into account for a reason. Namely, with only one of
them taken into account, some exact solutions (see, e.g.,
\cite{il1, we2, we3, iv1, iv2}) could be obtained (and that is
true not just for GB, but for any order of Lovelock corrections
and in any number of dimensions~\cite{prd}). Though when we take
into account different contributions, the equations become much
more complicated making it almost impossible to obtain any exact
solution in general case (for the first time solutions with both
terms taken into account were found in~\cite{deser}; in~\cite{we3}
a cosmological solution in exponential form was obtained). Yet,
the influence of the other (than leading) terms could be
significant -- for instance, in~\cite{we2}, while investigating a
(4+1)-dimensional model with matter in the form of a perfect fluid
in a ``pure'' Gauss-Bonnet case, we discovered an unusual behavior
in $w < 1/3$ case, and we believe it was caused by neglecting the
GR term in the equations of motion.

The (4+1)-dimensional case with both GR and GB contributions was studied in~\cite{ind}, but a full analysis of all regimes was not
performed there. In this paper we are going to do this -- we will give a description of all regimes, as well as try to
explain some of them analytically.

The second aim of this paper is to investigate the regimes in
presence of matter in the form of a perfect fluid. We reported
some features of the influence of matter in~\cite{we3}, but no
actual description was given. Namely, we reported that the
presence of matter leads to an increase of the ``probability'' of
smooth transition between the low- and high-energy Kasner regimes.
In the current paper we give a full description of all types of
transitions as well as describe the influence of the matter in
large. Finally, we generalize the results obtained -- both for the
vacuum and matter cases -- to higher even-dimensional models with
all possible Lovelock corrections taken into account.

This article is the logical continuation of~\cite{prd}. Indeed,
in~\cite{prd} we considered the flat anisotropic cosmological
models in the Lovelock gravity, derived the equations of motion
and investigated the dynamics with only the highest Lovelock
correction taken into account. In this paper we investigate the
lowest model ($D=4$, $n=2$) with all corrections taken into
account and based on the results predict the behavior of
even-dimensional models with all possible Lovelock corrections
considered.

\section{Equations of motion}

Equations of motion can be easily derived from the general ones, obtained in~\cite{prd}. In terms of Hubble functions they take a form:
dynamical equations

\begin{eqnarray}
\begin{array}{l}
2(\dot H_b+H_b^2)+2(\dot H_c+H_c^2)  +2(\dot
H_d+H_d^2)+2H_bH_c+2H_bH_d+2H_cH_d + \\
\\+ 8\alpha\Big [ (\dot H_b+H_b^2)H_cH_d +(\dot H_c+H_c^2)H_bH_d
+(\dot H_d+H_d^2)H_bH_c\Big ] + p=0
\end{array}
 \label{dyn}
\end{eqnarray}

\noindent (this is the first equation; the rest of them could be obtained via a cyclic indices transmutation) and a constraint equation

\begin{equation}
2H_aH_b+2H_aH_c+2H_aH_d+2H_bH_c+2H_bH_d+2H_cH_d+24\alpha
H_aH_bH_cH_d  =\rho.
 \label{constr}
 \end{equation}

\noindent In the vacuum case (\ref{dyn}) and (\ref{constr}) reduce
with $w=0$ and $\rho=0$. For simplicity we will refer to the model
with a non-zero matter contribution as the ``matter case'' (or the
``matter model'').

So in the matter case in addition to (\ref{dyn}) and
(\ref{constr}) one also needs an equation of state $p = w \rho$
for the perfect fluid and a continuity equation:

\begin{equation}\label{cont_eq}
\dot \rho + (\rho + p) (H_a + H_b +H_c +H_d) = 0.
\end{equation}

The system (\ref{dyn})--(\ref{constr}) in the vacuum and
(\ref{dyn})--(\ref{cont_eq}) in the matter case combined with the
initial conditions completely determines the evolution of the
Universe as a whole. Hence we are going to solve numerically these
equations to find out the future and past evolution of the model
with particular initial conditions. Scanning over the initial
conditions gives us the needed distribution of regimes.

\section{Vacuum model: special cases}

Before starting to present the results it would be an asset to
consider two special cases that can be found on our transition
maps. We call the first of them a 3-equal case with three Hubble
parameters equal to each other. In~\cite{ind} it was noted that
only this regime in a (4+1)-dimensional case has a smooth
transition between the low- and high-energy Kasner regimes. Using
$H_a = H_b = H_c = H$ and $H_d= h$ the dynamical equations reduce
to

\begin{equation}\label{3eqd1}
6H^2+4Hh+4\dot H+16\alpha Hh \dot H + 16\alpha{H}^{3}h++2\dot h+2{h}^{2}+8\alpha{H}^{2}{h}^2 + 8\alpha{H}^{2}\dot h = 0,
\end{equation}

\begin{equation}\label{3eqd2}
6\dot H + 12 H^2 + 24\alpha H^2 \dot H + 24\alpha H^4 = 0,
\end{equation}

\noindent and the constraint equation gives $h=-H/(1+4\alpha
H^2)$. Solving (\ref{3eqd1}) and (\ref{3eqd2}) gives

\begin{equation}\label{3eqdH}
\dot H = - \frac{2 H^2 (1+2\alpha H^2)}{1+4\alpha H^2},
\end{equation}

\begin{equation}\label{3eqdh}
\dot h = - \frac{2 H^2 (8\alpha^2 H^4 + 2\alpha H^2 - 1)}{(1+8\alpha H^2 + 16\alpha^2 H^4)(1+4\alpha H^2)}.
\end{equation}

Note that our equations are different from those in~\cite{ind},
and we claim ours to be correct\footnote{One can easily verify it
-- indeed, substituting $H_a = H_b = H_c = H$ into Eq. (7)
in~\cite{ind} (that is the dynamical equation that corresponds to
$H_d$, so it does not have $H_d$ itself but only $H_a$, $H_b$, and
$H_c$), one would directly obtain (\ref{3eqd2}); it is quite easy
to solve it to obtain (\ref{3eqdH}). The difference might be
caused by a wrong sign in $h=-H/(1+4\alpha H^2)$ (in~\cite{ind}
the authors have a plus instead of a minus).}. Still, the reason
why a smooth transition between the low- and high-energy Kasner
regimes occurs is the same -- the denominator of neither $\dot H$
nor $\dot h$ never crosses zero.

We call the other case a 2-equal -- now only two Hubble parameters
are equal to each other. Unlike the 3-equal case, here the
denominator of $\dot H_i$ can cross zero so a nonstandard
singularity could occur. But what is important, in this case there
also exists a smooth transition between the low- and high-energy
Kasner regimes, thus, the conditions for a smooth transition
between the low- and high-energy Kasner regimes are somehow
weakened. But since a 2-equal case still needs exact equality, the
measure of this transition is not improved.

\section{Vacuum model: results}

As we claimed above, we are going to make a scans over the initial
values of Hubble parameters to produce maps of trajectories.
Instead of producing 3D maps, as it was done in~\cite{ind}, we
will make several 2D plots with different values for the third
Hubble parameter. Thus the 3-equal case would be a point on our 2D
plot with coordinates that coincide with the fixed third Hubble
parameter, and the 2-equal case would be a diagonal line and two
perpendicular lines with the $x$- and $y$-coordinates equal to the
fixed third Hubble parameter. So we have three ``free'' Hubble
parameters, the fourth is calculated from the constraint equation
(\ref{constr}):

\begin{equation}\label{4th}
H_d^{(0)} = - \frac{H_a^{(0)} H_b^{(0)} + H_a^{(0)} H_c^{(0)} + H_b^{(0)} H_c^{(0)}}{H_a^{(0)} +H_b^{(0)} +H_c^{(0)} + 12\alpha H_a^{(0)}
H_b^{(0)} H_c^{(0)}}.
\end{equation}

Now let us have a look on the denominator of (\ref{4th}). Later,
when dealing with the matter case, we will see that this
denominator (more specifically, its sign) is an essential thing in
describing the matter regimes. For now let us note that if all
three Hubble parameters are positive, the denominator is also
positive, but if one of the three Hubble parameters is negative,
the denominator could be (but not always) negative. We refer to
the ($H_a^{(0)}$, $H_b^{(0)}$, $H_c^{(0)}$) triplet as a
``positive'' one if it produces the positive denominator of
(\ref{4th}) and as ``negative'' if it is negative.

Thus we fix the initial value for one of the Hubble parameters
(without loss of generality, let it be $H_a^{(0)}$) and scan over
$H_b^{(0)}$ and $H_c^{(0)}$ in [0\dots 1.5$\div$ 2] range. In
previous studies~\cite{we2, we3, we1} we usually considered
\mbox{[0, 1]} range, but now, for the demonstrative reasons we
decided to expand it. As for $H_a^{(0)}$, we vary it roughly in
[$-$1.5 \dots 1.5] range. Also, as in all previous studies, we
consider only the initially-expanding Universe: $\sum_i H_i^{(0)}
\geqslant 0$. First, we present our results for the positive
$H_a^{(0)}$, then for the negative.

Before giving the results, let us summarize what we would expect.
For the past evolution, we expect either a standard singularity
(high-energy Kasner regime) or a nonstandard one. For the future,
there are three possibilities: a low-energy Kasner regime, a
recollapse or a nonstandard singularity. For simplicity we denoted
them as summarized in Table. With the sign of triplet taken into
account, we denote the trajectories like their types with
appropriate signs, e.g., type VI+ corresponds to type VI with a
positive triplet.

\begin{table*}
{{\bf Table.} The classification of possible trajectories in the vacuum (4+1)-dimensional GR+GB model}\\
\begin{tabular}{c|c|c}
\hline
From & To & Design. \\
\hline

Standard singularity & Kasner & I \\
 & Recollapse & II \\
 & Non-standard sing. & III \\
\hline

Non-standard. sing. & Kasner & IV \\
 & Recollapse & V \\
 & Non-standard sing. & VI \\
\hline

\end{tabular}
\end{table*}

\begin{figure}
\includegraphics[width=0.8\textwidth]{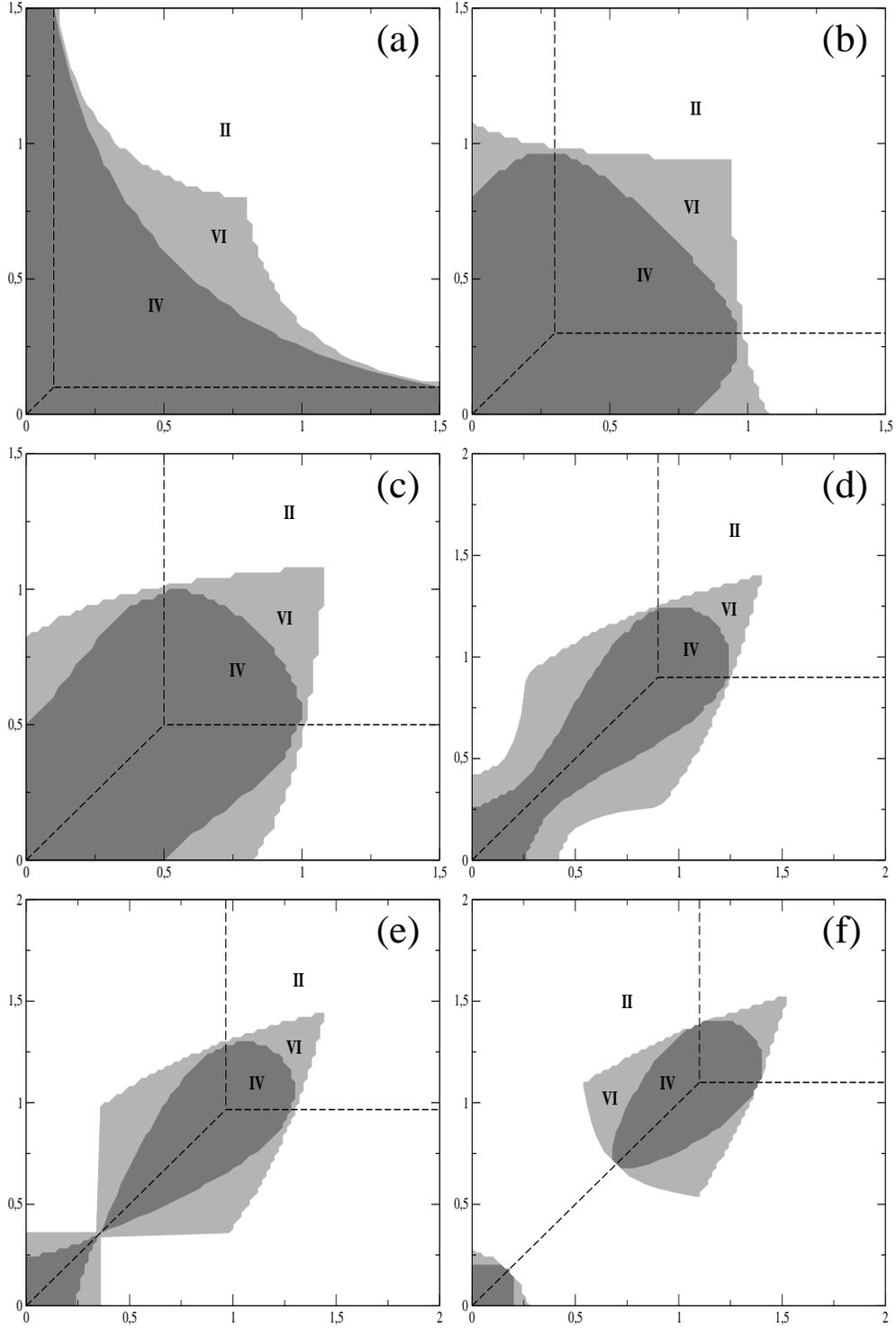}
\caption{ Transitions for $H_a^{(0)} > 0$ in the vacuum case. The
$H_a^{(0)}$ value is increasing from (a) to (f). Trajectory types
are denoted according to the Table. Dashed lines correspond to the
type-I trajectories. }\label{vac1}
\end{figure}

The results for the \underline{$H_a^{(0)} > 0$ case} are given in
Fig. \ref{vac1}. In there, we plotted the resulting past and
future behavior for the fixed $H_a^{(0)}$, and with $H_b^{(0)}$
vs. $H_c^{(0)}$ as coordinates. The resulting transitions are
denoted according to the Table, the dashed lines are type I
trajectories. The value for $H_a^{(0)}$ is increasing from Fig.
\ref{vac1}(a) to Fig. \ref{vac1}(f): $H_a^{(0)} = 0.1$ in (a),
$H_a^{(0)} = 0.3$ in (b), $H_a^{(0)} = 0.5$ in (c), $H_a^{(0)} =
0.9$ in (d), $H_a^{(0)} = 0.966$ in (e) and $H_a^{(0)} = 1.1$ in
(f).

From Fig. \ref{vac1} one can understand how the transitions evolve
with changing $H_a^{(0)}$. At some value -- in our case with
$\alpha=1$ it occured at $H_a^{(0)} \approx 0.966$ -- two type-IV
regions ``detach'' from each other; with further growth of
$H_a^{(0)}$ the bottom-left region shrinks; the upper-right one
also decreases in size and moves by diagonal towards the growth of
$H_c^{(0)}$ and $H_b^{(0)}$. The $H_a^{(0)}$ value this
``detachment'' occurs at is dependent on $\alpha$ and governed by
a high-order equation, so it is impossible to give it an exact
analytical expression.

\begin{figure}
\includegraphics[width=1.0\textwidth]{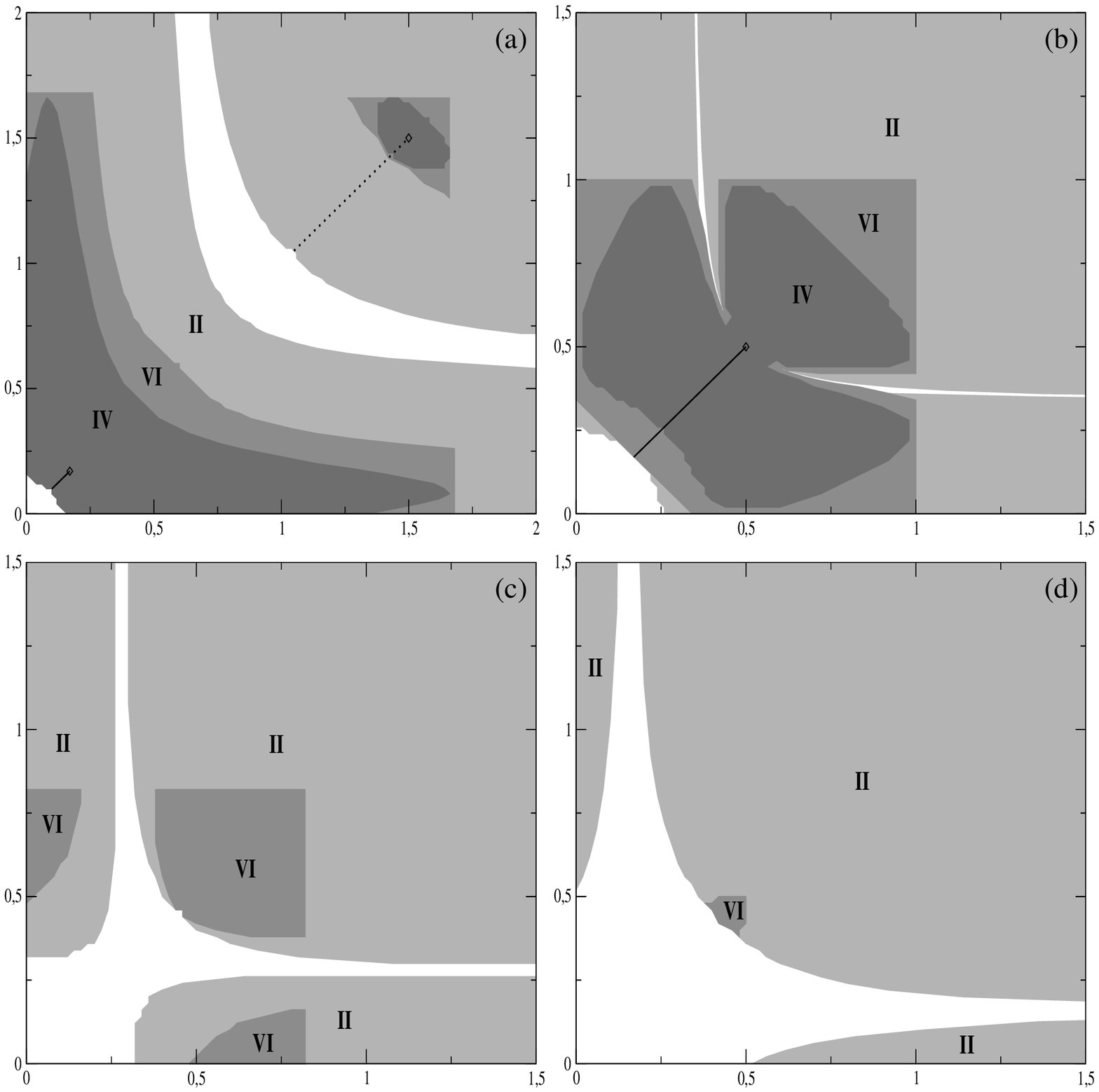}
\caption{ Transitions for $H_a^{(0)} < 0$ in the vacuum case. The
$H_a^{(0)}$ value is decreasing ($|H_a^{(0)}|$ is increasing) from
(a) to (d).Trajectory types are denoted according to the Table.
Solid lines correspond to the I+ trajectories, dotted -- to the
I$-$. }\label{neg}
\end{figure}


The \underline{$H_a^{(0)} < 0$ case} is more complicated than the
previous one. Indeed, with all three ``free'' initial Hubble
parameters being positive, the last Hubble parameter is always
negative while \mbox{$\sum_i H_i^{(0)} > 0$} (one can simply
verify this from (\ref{4th})). But with $H_a^{(0)}$ being
negative\footnote{The same with the previous subsection, we fix
$H_a^{(0)}$ and plot 2D figs with $H_b^{(0)}$ and $H_c^{(0)}$ as
the $x$- and $y$-coordinates.} the situation becomes more
complicated. First of all, the calculated Hubble parameter could
be either positive or negative. First case is partially
epistemologically unimportant to us -- this case is equivalent to
$H_a^{(0)} > 0$ and thus it does not bring us new regimes, but for
the completeness purposes we keep it. The second case could bring
new regimes -- indeed, with one of the three ``free'' Hubble
parameters being negative, the denominator of (\ref{4th}) could be
negative and its regimes, as we will see in the case with a matter
source, are distinct from the positive triplets case. However, two
negative initial Hubble parameters could also lead to $\sum_i
H_i^{(0)} < 0$.

The results are presented in Fig. \ref{neg}. Therein we gave 2D
scans for four different values of $H_a^{(0)}$: $H_a^{(0)} =
-0.15$ in (a) panel, $H_a^{(0)} = -0.25$ in (b), $H_a^{(0)} =
-0.3$ in (c) and $H_a^{(0)} = -0.5$ in (d). At $H_a^{(0)} \lesssim
-0.5$ type-VI trajectories completely ``disappear'' from the
scene, hence we decided to skip the plots of that kind. A white
region corresponds to the initial conditions with $\sum_i
H_i^{(0)} < 0$; hyperbola-like regions originate from zeros of the
denominator in the $H_d^{(0)}$ expression:
$H_a^{(0)}+H_b^{(0)}+H_c^{(0)}+12\alpha H_a^{(0)} H_b^{(0)}
H_c^{(0)} = 0$. Thus, the upper-right ``half'' of the 2D plots in
Fig. \ref{neg} has a negative denominator of $H_d^{(0)}$
(\ref{4th}) while the others have positive denominators.

In addition to the type II, IV, and VI trajectories at low values
of $|H_a^{(0)}|$, one can find type-I trajectories as well. There
as well exists the transition of I$-$ type, that we can call ``the
rarest'' trajectory. Indeed, it exists only at $H_a^{(0)} < 0$,
only at 2-equal lines, at low enough $|H_a^{(0)}|$ (type-IV should
not disappear yet -- $H_a^{(0)} > -1/(4\sqrt{\alpha})$), and on
the upper ``half'' of the plot; the I$-$ trajectories are denoted
by a dotted line in Fig. \ref{neg}(a). The I+ transitions are
denoted as solid lines in Fig. \ref{neg}(a) and (b). One can see
that type-I trajectories exist between the lower boundary of the
$\sum_i H_i^{(0)} > 0$ and some point that we denoted as diamonds
in Fig. \ref{neg}(a) and (b). These points correspond to the
3-equal situation: indeed, from Sec. III we remember that in the
3-equal regime the calculated Hubble parameter is always negative.
If we assume $H_a^{(0)}$ to be that parameter, then the 3-equal
situation occurs at

\begin{equation}\label{3eq}
H_b^{(0)} = H_c^{(0)} = H_d^{(0)} = -\frac{1}{8}\,\frac{1\pm\sqrt{1-16\alpha {\displaystyle (H_a^{(0)})^2}}}{\alpha H_a^{(0)}}.
\end{equation}

\noindent Two different 3-situations are infinitely separated at
$H_a^{(0)}\to (0-0)$, become closer with the growth of
$|H_a^{(0)}|$ and coincide when $H_a^{(0)} = -1/(4\sqrt{\alpha})$
(with $\alpha=1$ it happened at $H_a^{(0)} = -0.25$; the situation
is presented in Fig. \ref{neg}(b)).

\section{Vacuum model: discussion}

In the first half of the article we dealt with the vacuum
(4+1)-dimensional flat anisotropic model in the
Einstein-Gauss-Bonnet gravity. One can see that the type-II
trajectories are dominating in both $H_a^{(0)} > 0$ and $H_a^{(0)}
< 0$ situations; in other words, the type-II trajectories are
dominating all other regimes in the model considered. In the
$H_a^{(0)} > 0$ case we have IV and VI-type trajectories more or
less abundant, their presence is in general lessened with the
growth of $H_a^{(0)}$, but they never vanish. In the $H_a^{(0)} <
0$ case the situation is more dramatic: with small $|H_a|$ we have
I, II, IV, and VI type trajectories, but with the growth of
$|H_a|$ we first ``loose'' I and IV, then VI, therefore finally at
$H_a^{(0)} \lesssim -0.5$ (with $\alpha = 1$; with a different
value for $\alpha$ this will happen at another value for the
$H_a^{(0)}$) we have only type-II transitions.

\begin{figure}
\includegraphics[width=1.0\textwidth]{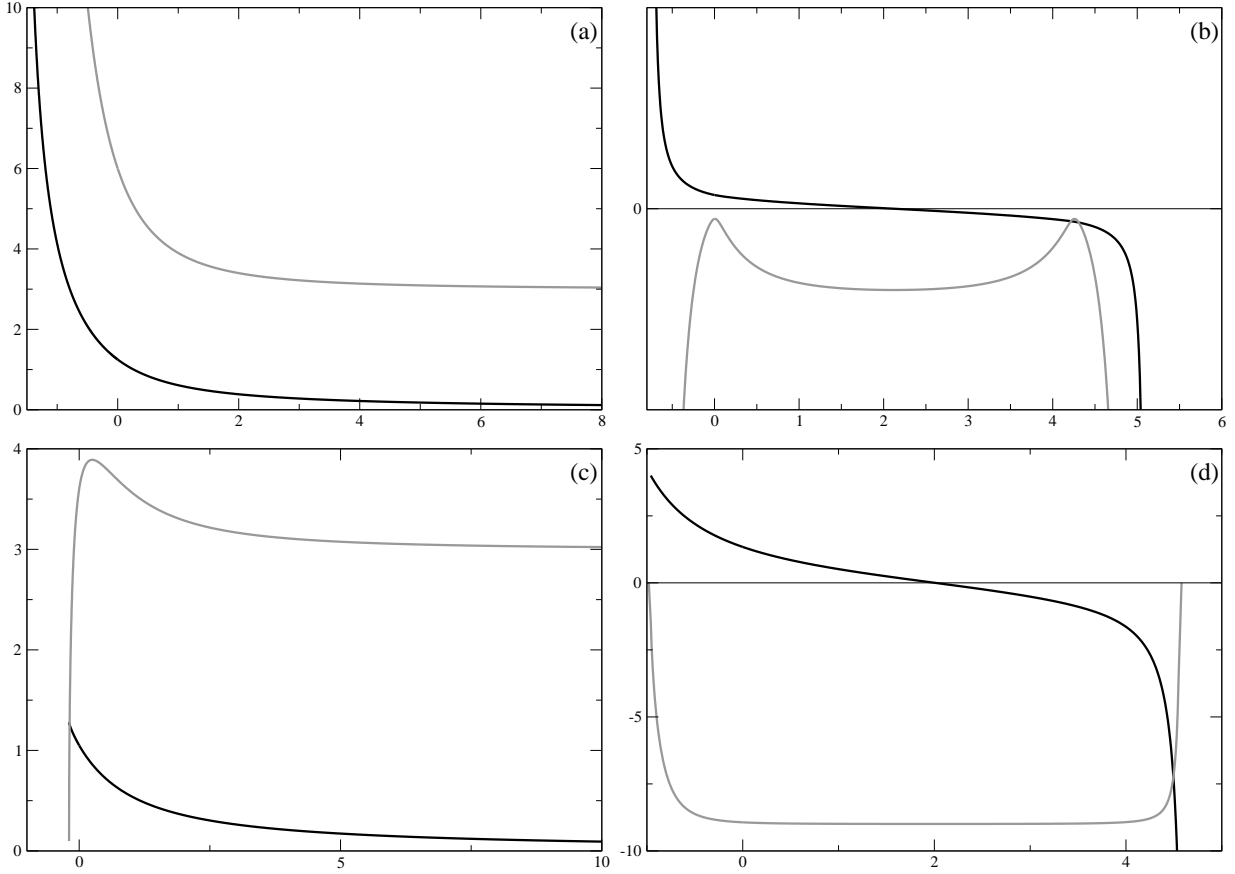}
\caption{Examples of different type of transitions. Black curve is
the expansion rate in terms of the Hubble parameters $\sum_i H_i$,
grey curve is the value for the denominator in the $\dot H_i$
expression. Type I is in (a), II -- (b), \mbox{IV -- (c)}, and VI
-- (d).}\label{tr_ex}
\end{figure}

In Fig. \ref{tr_ex} we presented the examples of all the possible
transitions. Black curves correspond to the expansion rate in
terms of the Hubble parameters $\sum_i H_i$, and grey curves
represent the value for the denominator of the $\dot H_i$
expression. Time is along $x$-axis and is normalized in a way to
set $\alpha=1$; zero-point $t=0$ corresponds to the start of
integration. Type I is in (a), II -- (b), IV -- (c), and VI --
(d). Hence there are no type-III and V transitions, and now we
will try to explain why. One can see that the nonstandard
singularity occurs only in the vicinity of the standard one (but
it is not necessary that it will happen -- see Fig. \ref{tr_ex}(a)
for example). From Figs. \ref{tr_ex}(a) and (c) one can learn that
the trajectories, starting from the positive values of the
denominator end up in the Kasner regime. We checked that for the
scans in \mbox{Fig. \ref{vac1}}, and found that the type-IV regime
completely coincides with an initially positive denominator. The
reason behind this link between the positivity of denominator and
the late-time Kasner behavior is a bit aside from the goals of
this paper and is yet to uncover -- we are going to devote a
separate paper to the nonstandard singularity in flat anisotropic
models in the Lovelock gravity. Anyway, type III and V
trajectories cannot originate from the positive initial value of
the denominator. As for the negative value, the remaining two
possibilities are presented in Figs. \ref{tr_ex}(b) and (d). One
could think of the possibility for the type-III trajectory to
occur on the singular analog of the left branch in Fig.
\ref{tr_ex}(b) -- on the short ``window'' between the standard and
nonstandard singularities, but our numerical analysis demonstrated
that it is not happening. We believe this situation could not even
be constructed: crossing the nonstandard singularity changes the
sign of the denominator leaving the numerator unchanged; thus at
this point the sign of $\dot H_i$ changes, making it useless to
try to construct any continuation beyond the nonstandard
singularity.

\section{Model with matter}

In this section we start our investigation of the models filled
with matter in the form of a perfect fluid. In this case the
constraint equation gives us a bit different expression for the
4th Hubble parameter:

\begin{equation}
H_d^{(0)} = - \frac{ 2H_a^{(0)}H_b^{(0)} + 2H_a^{(0)}H_c^{(0)} + 2H_b^{(0)}H_c^{(0)}  - \rho_0 }
{2H_a^{(0)} + 2H_b^{(0)} + 2H_c^{(0)} + 24\alpha H_a^{(0)}H_b^{(0)}H_c^{(0)}}.
\label{4hubble}
\end{equation}

From (\ref{4hubble}) one can easily see that the expression for
$H_d^{(0)}$ is split into two parts -- $\rho$-dependent and
$\rho$-independent. The latter remains the same with varying
$\rho$ while the former changes the value for $H_d^{(0)}$. At this
point the positive and negative triplets start to play an
important role. These two have a huge difference -- the positive
triplets have no boundary on the density while the negative do.
Indeed, keeping in mind that $\sum H_i^{(0)}\geqslant 0$ and the
fact that the increasing density decreases $H_d^{(0)}$, even if at
$\rho=0$ we have $\sum H_i^{(0)} > 0$, at some value for $\rho$ we
will have $\sum H_i^{(0)}= 0$. This difference between the
positive and negative triplets affects the details of their
regimes, so while describing the regimes we will describe them
separately.

\section{Model with matter: results}

Now let us describe the matter regimes. The first one to describe
is \underline{I+ case}. In the vacuum case, only I+ and I$-$ have
the smooth transition from the standard singularity to the Kasner
expansion, but, as we noted in~\cite{we3}, in the model with
matter it is no longer the case. In the presence of matter the
transition remains unchanged: it is from the standard singularity
to the expansion. The difference between the $w<1/3$ and $w>1/3$
cases lies in the isotropisation: at $w > 1/3$ the initial
singularity becomes isotropic (and the expansion is the
GR-dominated Kasner) while at $w < 1/3$ we have a standard
GB-dominated singularity and an isotropic expansion.

\begin{figure}
\includegraphics[width=1.0\textwidth]{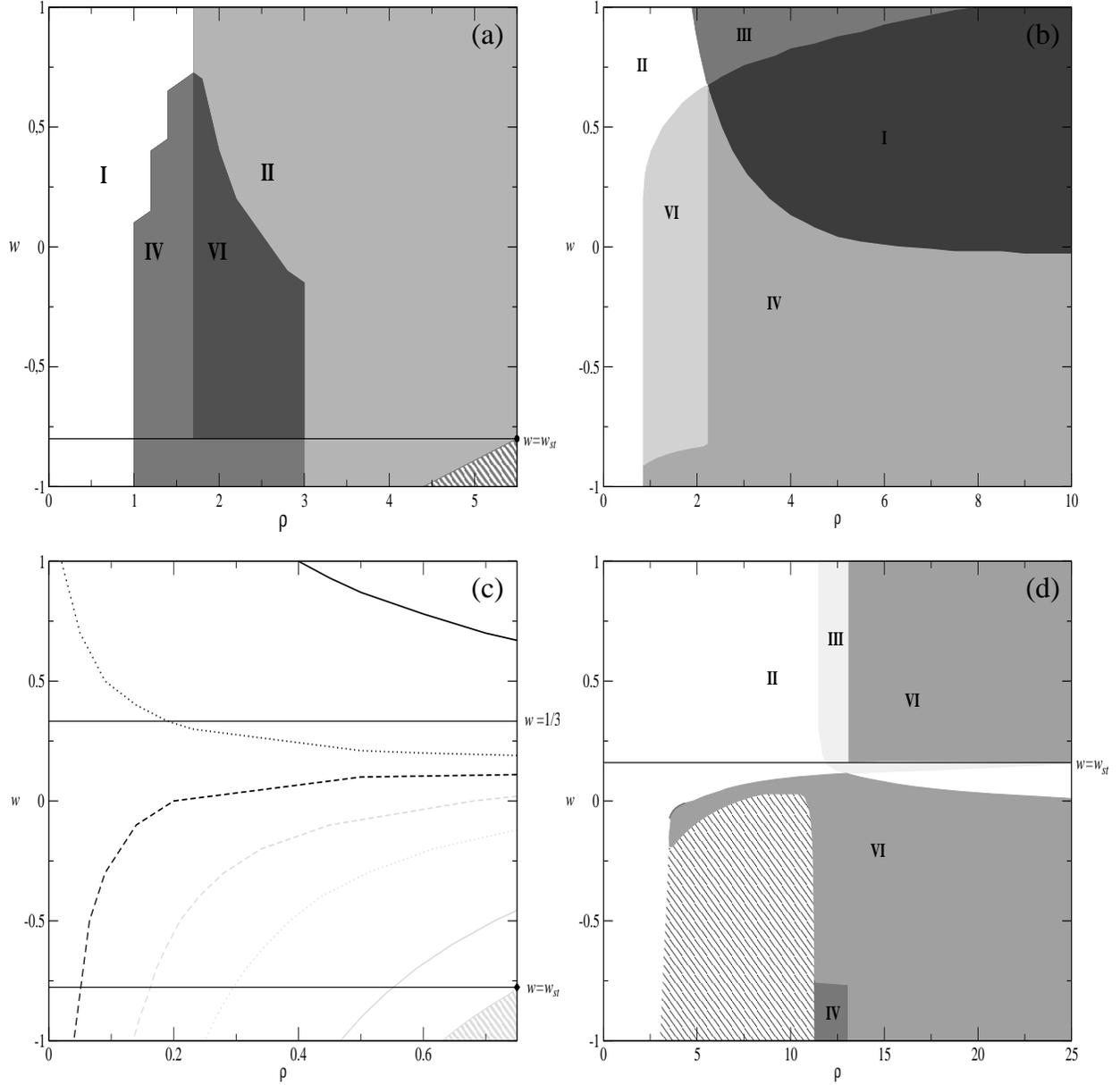}
\caption{An example trajectory map in $(\rho, w)$ coordinates for
I$-$ (a), II+ (b) and II$-$: $w_{st} < 0$ (c) and $1/3 > w_{st} >
0$ (d) subcases. The stroked region in (a), (c), and (d) panels
corresponds to the periodical trajectories reported in~\cite{we3}.
See the text for details. }\label{first41}
\end{figure}

Trajectories of \underline{I$-$ type}, as we already mentioned,
are ``the rarest'' trajectories. In the presence of matter the
regimes changes according to Fig. \ref{first41}(a); the trajectory
types are denoted in the figure. As a regime with the negative
triplet, it has the maximal value for density; the $x$-axis of the
Fig. \ref{first41}(a) covers all possible values for density.

The transitions in the $(\rho, w)$ map for \underline{II+ type}
trajectory are given in Fig. \ref{first41}(b). There is a
difference between the regimes at $w<1/3$ and $w>1/3$: in the
$w<1/3$ case, the GR-Kasner ``transforms'' into an isotropic
expansion while in the $w>1/3$ case, the isotropic one is the
initial singularity.

From Fig. \ref{first41}(b) one can easily see what brings an
addition of matter to this regime. First of all, new, unknown for
the vacuum case, type-III regime is introduced. It doesn't cover a
wide range of $\rho$ and $w$, but its presence is still important.
Secondly, the type-I trajectory in this case is unbounded from
high densities. Thus, we find that the presence of matter in the
form of a perfect fluid makes the type-I transition common in
terms of initial parameters in contrast with the vacuum case,
where one requires fine-tuning to achieve this regime. One can
also notice that the type-I transition covers both $w>1/3$ and
$w<1/3$ cases, so the expansion can be both Kasner-type and
isotropic, respectively. Finally, the region of type-IV
trajectories is also unbounded from high densities and lies mostly
in the lower half of the $(\rho, w)$ map; later we will see that
this is more or less typical for the type-IV trajectories.

The \underline{II$-$ case} is more complicated than the others.
Namely, it has a different behavior in the $w_{st} < 0$ and $1/3 >
w_{st}
> 0$ cases. We consider them separately, and now we only note one
feature of the $w>1/3$ regimes: the initial singularity is
GR-dominated in contrast with I+ and II+ where in the $w>1/3$ case
the initial singularity was isotropic.

In \underline{$w_{st} < 0$ case} almost all the trajectories
remain the same -- II type. Since this is ``standard singularity
$\to$ recollapse'', one can find its ``lifetime''. To demonstrate
the influence of matter on the dynamics, we plotted the resulting
lifetimes on the $(\rho, w)$ map (see Fig. \ref{first41}(c)). The
stroked region in the bottom-right corner marks the positions of
periodic trajectories -- a new type of solution in the vicinity of
the stationary one ($\rho = \rho_{st},~w=w_{st}$), described
in~\cite{we3}. Also one can notice that there is no abrupt
difference between the $w<1/3$ and $w>1/3$ regimes.

And one last point regarding this regime -- in the case of
$w_{st}<-1$ (this is possible in some triplets) we do not have
anymore periodical trajectories (and so only type II trajectories
remains) as they are all located at $w < -1$ which is beyond
physical assumptions.

\underline{The case with $1/3 > w_{st} > 0$} is different from the
described above $w_{st} < 0$ -- it has more different regimes. In
Fig. \ref{first41}(d) we presented the map of transitions in
$(\rho, w)$ coordinates.

\begin{figure}
\includegraphics[width=1.0\textwidth]{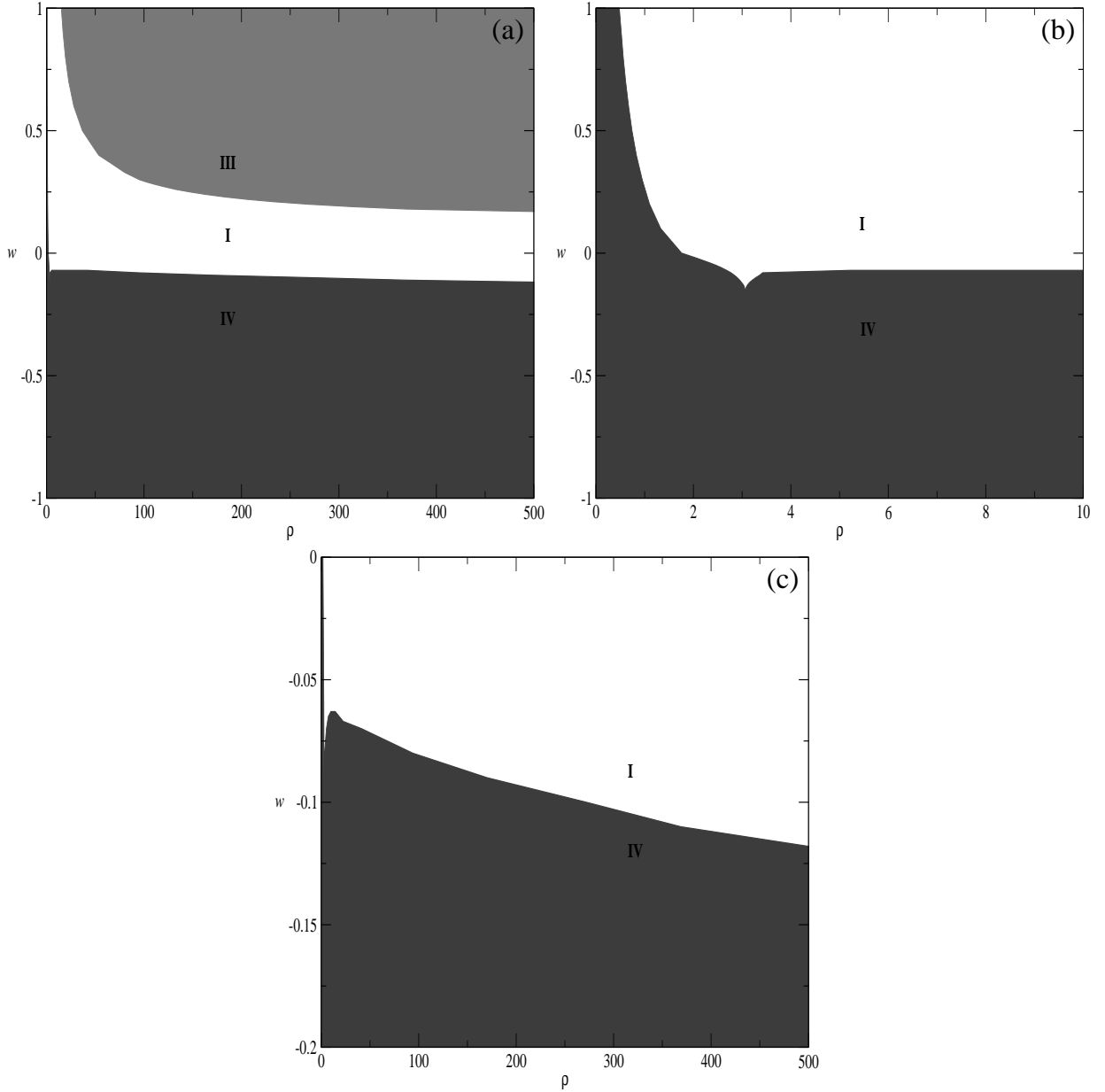}
\caption{An example trajectory map in the $(\rho, w)$ coordinates
for the IV+ case. In (a) we can see the large-scale structure of
the transitions (up to very large values of $\rho$), in (b) we
presented the boundary between I and IV at low densities with
feature, and finally in (c) -- the boundary between I and IV at
large.}\label{4plus}
\end{figure}

The map of trajectories for the \underline{IV+ case} is given in
Fig. \ref{4plus}. There in panel (a) we present a large-scale
structure of the transitions; in panel (b) -- a detailed structure
of the boundary between I and IV at low densities and in panel (c)
-- the boundary between I and IV at large.

\begin{figure}
\includegraphics[width=1.0\textwidth]{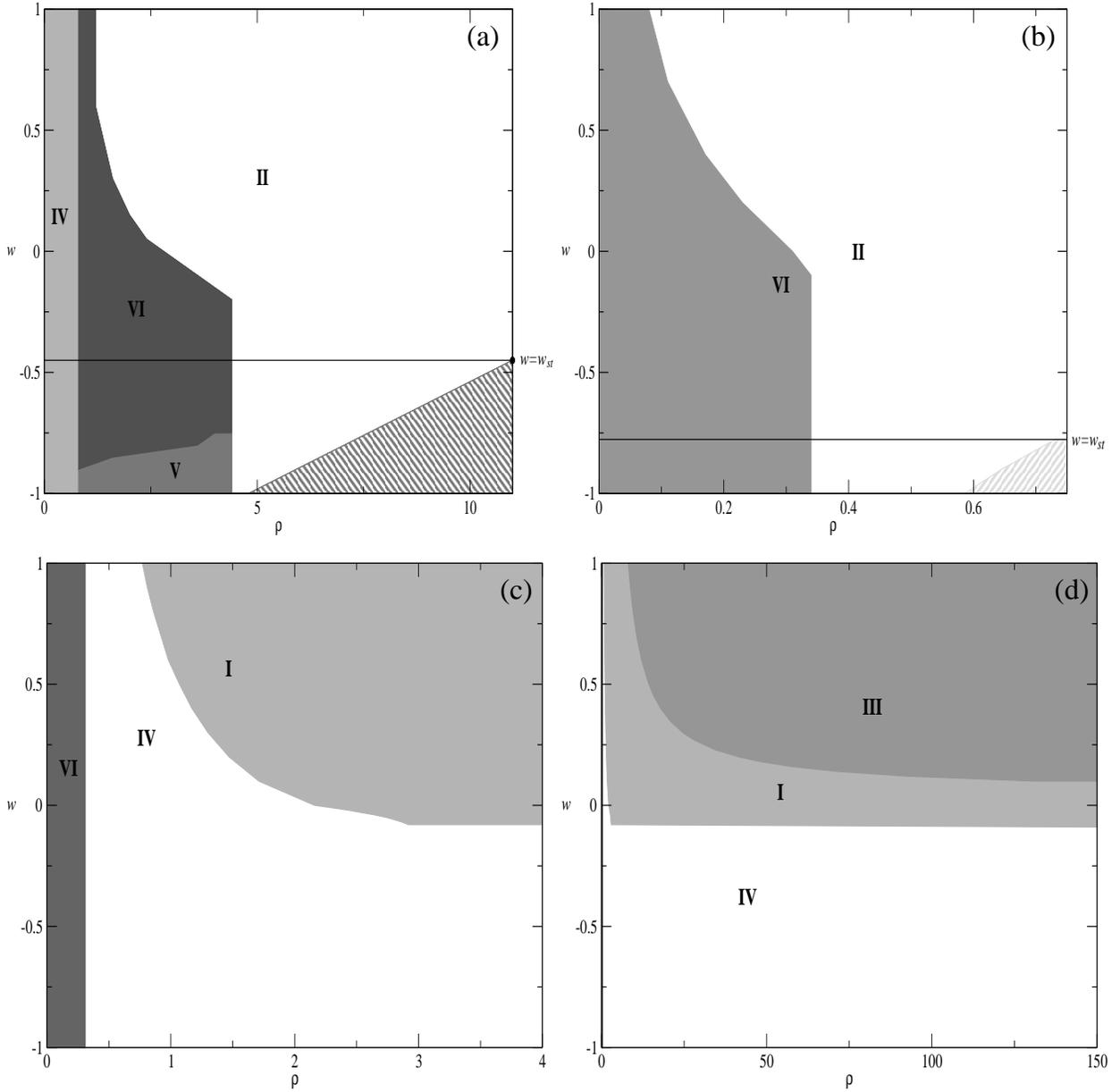}
\caption{ (a) Example of the $(\rho, w)$ transition map for the
IV$-$ case; (b) -- the same but for the VI$-$ case. One can see
the region with periodical trajectories in the bottom-right corner
of panels (a) and (b). In (c) and (d) panels -- $(\rho, w)$
transition map for the VI+ case: structure at low $\rho$ in (c)
and large-scale structure in (d). }\label{last32}
\end{figure}

The resulting $(\rho, w)$ transition map for the \underline{IV$-$
case} is given in Fig. \ref{last32}(a). The type IV trajectories
are quite bounded in there, but we have another regime, unseen in
the vacuum case -- type V. With it we have all the six possible
regimes in the presence of matter. With high enough $w_{st}$ we
have a relatively large area with periodic trajectories.

Two final regimes, \underline{VI+} and \underline{VI$-$}, are
presented in Figs. \ref{last32}(b), (c), and (d). In Fig.
\ref{last32}(b) we presented type VI$-$ and in Figs.
\ref{last32}(c) and (d) -- type VI+: the behavior at low values
for $\rho$ in (c) and the large-scale structure of the transitions
in (d).

\section{Model with matter: discussion}

\begin{figure}
\includegraphics[width=1.0\textwidth]{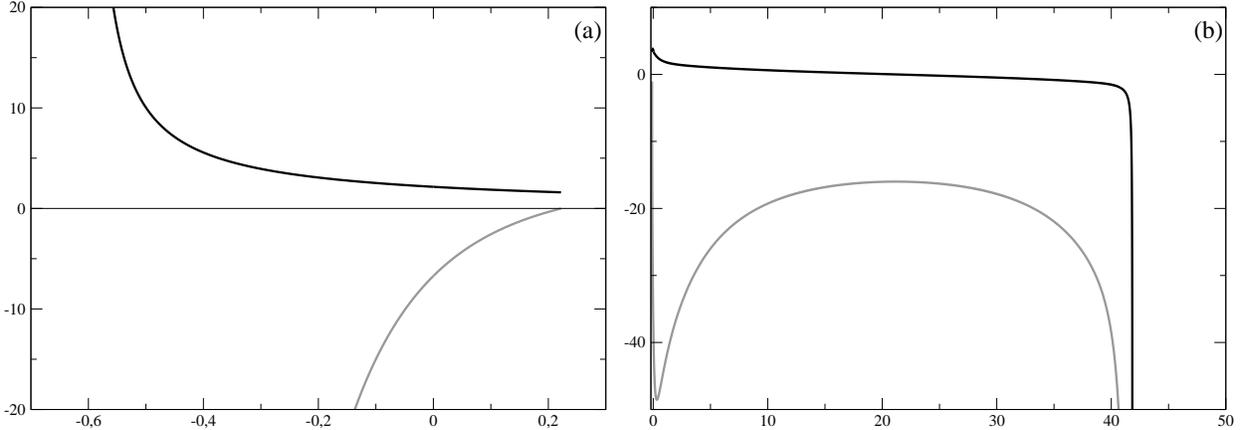}
\caption{Type-III (a) and type-V (b) transitions in the case with
matter. Black line represent $\sum H_i$, grey -- denominator of
$\dot H$ expression.} \label{IIImatterV}
\end{figure}

In the second part of the paper we studied what influence the
matter in the form of a perfect fluid exerts upon the dynamics of
flat (4+1)-dimensional anisotropic models in the
Einstein-Gauss-Bonnet gravity. One of the two main changes from
the vacuum case is a substantial increase of the type-I
transitions we previously reported in~\cite{we3}. An interesting
feature -- only positive triads have type-I transition in the
presence of matter, negative (with an exception of I$-$) do not.
In this paper we demonstrate this increase ``quantitatively''.
Another feature is the appearance of type-III and type-V
trajectories. First of them appeared exactly the way we discussed
in the vacuum case: on the singular analog of the left branch of
Fig. \ref{tr_ex}(b). We presented both III and V in Fig.
\ref{IIImatterV}\footnote{Remarks to Fig. \ref{tr_ex} regarding
the time are valid to Fig. \ref{IIImatterV} as well.}. Thus, the
presence of matter changes the dynamics to make such trajectories
to exist.

One can find different ``features'' on the $(\rho, w)$ transition
maps and most of them (those that are unrelated to $w_{st}$) are
just triplet-specific features. A feature at the II+ transition
map (Fig. \ref{first41}(b)) where all five regimes ``touch'' is a
good example. If we draw a $w={\rm const}$ line from that point to
$\rho=0$ we will notice that the lifetimes for type-II
trajectories along this line are the same and equal to that of the
vacuum case. It is obvious that the lifetime in the vacuum case is
triplet-specific and varies from triplet to triplet, hence the
location of features of that type also varies from triplet to
triplet. The same situation is with the features of other kinds.

Now we want to draw your attention to the II$-$ $1/3 > w_{st}$
case. One can see that it is very different from other cases. It
has periodical trajectories at an ``unusual'' (from the other
regimes that have periodical trajectories) place and it has much
more type-VI transitions than the other regimes do. The reasons
behind this behavior are not clear; probably, that is due to
crossing the ``special solution'' $w=0$, found in~\cite{we2}
numerically for the GB+matter case (later in~\cite{prd} we
demonstrated that this is a general feature for all
even-dimensional flat anisotropic models in the Lovelock gravity
with only the higher-order correction taken into account).

One can notice that generally, inside a triplet, the transition
maps have a lot in common. For the positive triplet we have
III$\to$I$\to$IV ``regimes stratification'' (over $w$) on large
$\rho$: in the II+ case (Fig. \ref{first41}(b)), we have the same
situation in IV+ (Fig. \ref{4plus}) and VI+ (Fig. \ref{last32}(c,
d)). In the II+ case type-III trajectory ``disappears'' at some
density, but that is just an effect of a particular triplet: in
another triplet with a lower value for $w$ at which the feature
occurs we will have all the three regimes at high densities.
Negative triplets also have some features in common. They all have
type-II at the intermediate and high values of density and, if
$w_{st}
> -1$, they have periodical trajectories. They all as well (except
II$-$) have type-VI at low--intermediate values of density.

Therefore we see that the positive triplets have a high-density
regime stratification over $w$ while the negative triplets have it
over the density itself. Keeping in mind the difference between
the triplets one can explain this: for negative triplets the
density is more important -- high density can cause $\sum_i
H_i^{(0)} < 0$, so the value for density determines initial
regimes. For positive triplets all values of density are
acceptable hence the role of density is less important and the
equation of state determines the regimes.

\section{Higher-dimensional cases}

Finally, as we claimed at the beginning, we are going to
generalize the obtained results on the higher-dimensional flat
anisotropic models in the Lovelock gravity. First of all,
in~\cite{prd} we obtained the equations of motion for the flat
anisotropic cosmological model in the Lovelock gravity in any
number of dimensions with only the highest possible correction
taken into account. We also demonstrated how one can obtain the
equations for the mixture of corrections. If we take into account
all possible corrections, the resulting equations for different
orders would have a similar structure, and so would the solutions.
Therefore we expect the II-type trajectories to dominate in all
the even-dimensional models. The negative multiplets (the analogs
of triplets from (4+1)) would as well  have a more complicated
structure than that in Fig. \ref{neg}. This is caused by more
possible zeros of the denominator of the calculated Hubble
parameter. In its turn, it will increase the effective presence of
type-I, IV, and VI trajectories. The gain in type-I is also due to
the increase of possible ($D-k$)-equalities, but still the measure
will be incomparable with other types of regimes.

We do not expect any changes in the negative multiplets -- they
are governed by the density -- but there might be changes in the
positive multiplets. Indeed, they are governed by the equation of
state, and its influence is dependent on the number of dimensions.
The equation of state that corresponds to the highest possible
contribution ($n$-order) is $w_{eq} = 1/(2n-1)$. With the growth
of $n$ we have $w_{eq}\to (0+0)$; as one of the consequences one
would expect a decrease of the presence of the II$-$ type
trajectory with $0 < w_{st} < w_{eq}$ -- the corresponding range
of $w$ would shrink eventually decreasing the presence. As another
consequence one would expect a decrease of the type-I presence in
the positive multiplets (apart from I$-$ we do not have type-I in
the negative triplets in (4+1)). This is caused by the same
``shrinking'' $w_{eq}\to (0+0)$ while $w=0$ is the ``special
solution'' for the even-dimensional case~\cite{prd}. Thus the
region $0 < w < 1/3$ from (4+1) (where the type-I trajectories are
located) with the growth of $n$ would shrink resulting in a
decrease of presence  of the  type-I transitions.

Also, the characteristic scale of the Hubble parameters where
types I, IV, and VI occur will drop with the growth of $n$. This
is due solely to the number of dimensions -- indeed, the same
amount of matter produces lower density in the higher dimensions.
Our calculations for $D=6$ and $D=8$ confirmed that. Thus, purely
due to the growth of $n$ the effective presence of the type-I, IV,
and VI trajectories is decreasing.

\section{Conclusions}

In this paper we gave the first complete description of all the
regimes in the (4+1)-dimensional flat anisotropic cosmological
model in the Einstein-Gauss-Bonnet gravity. All through the paper
we hold the condition on the initial values of Hubble parameters:
$\sum H_i^{(0)} > 0$. With the $\sum H_i^{(0)} < 0$ case taken
into account, our description will be complete. And the $\sum
H_i^{(0)} < 0$ case could be easily obtained from the $\sum
H_i^{(0)} > 0$ case: indeed, assuming $\sum H_i^{(0)} < 0$ we
effectively only make a $t\to -t$ transform, so we will have the
same, but time-reversed results, such as, ``standard singularity
$\to$ Kasner'' turns into ``Kasner $\to$ recollapse'' and so on.
With this remark in mind, our description of the regimes is
complete for all the possible values of the initial Hubble
parameters.

We have found that the type-II transition (``standard singularity
$\to$ recollapse'') dominates over all the other types of
trajectories in the vacuum case, and describe it both
qualitatively and quantitatively. We as well found that the type-I
transition (``standard singularity $\to$ Kasner regime'') is
spread more widely than it was assumed~\cite{ind}, but still not
enough to compete even with type IV.

If we add matter in the form of a perfect fluid, the situation
changes drastically. Not only the abundance of the type-I
transition becomes comparable with abundances of other
types~\cite{we3}, but we also obtain new regimes that are unknown
for the vacuum case\footnote{Let us note -- considering the $\sum
H_i^{(0)} < 0$ case would not bring these two cases for they are
time-reversal to each other.} (type-III and V). With these two, in
the matter case we have all the possible transitions between two
possibilities for the past (standard and nonstandard
singularities) and three for the future (the Kasner regime,
recollapse and nonstandard singularity) evolutions. We gave the
description of the influence of matter on all the initially
possible vacuum regimes.

Apart from the description of the matter regimes, we found that at
intermediate--high densities all the regimes (inside their
triplet) look alike: positive triplets have III$\to$I$\to$IV
``regime stratification'' (with a decrease of $w$) while negative
ones have VI$\to$II ``regime sequence'' (with an increase of
$\rho$). Thus, the influence of matter is stronger than we
anticipated: it does not just significantly increase the abundance
of the type-I transition and adds the two remaining possible
trajectories, but also almost ``erases the bounds'' between
different trajectories inside the triplet. But there are
exceptions like II$-$ with $0 < w_{st} < 1/3$ that we discussed in
the relevant section.

Finally we generalized the obtained results on the
higher-dimensional case. We expect that the vacuum
even-dimensional flat anisotropic models in the Lovelock gravity
of the $n$ order are dominated by the type-II transition. The
presence of the other types of transitions is in general more and
more suppressed with the growth of $n$. The presence of matter
acts the same (as described for (4+1)) way in the case of negative
multiplets and the same but with decreases of the presence of
type-I trajectories with the growth of $n$ in the case of positive
multiplets. Therefore, the even-dimensional vacuum models are
degenerative from the point of view of the presence of a smooth
transition between the standard singularity and the cosmological
expansion; the presence of matter in the form of a perfect fluid
lifts this degeneration but with the growth of $n$ this
improvement decreases. Let us note, though, that ($2+1$) in $L_1$
differs from the described above picture~\cite{2plus1einst};
hence, the analysis is valid only a t $n\geqslant 2$.

\end{document}